\documentclass[prd,twocolumn]{revtex4}

\usepackage{graphicx}
\usepackage{amsmath}
\usepackage{multirow}
\usepackage{hyperref}
\usepackage{multimedia}

\newcommand{\BibUrl}{\url}

\begin{document}

\title{On the gravitational origin of the Pioneer Anomaly}

\author{Siutsou I. A.}
\email{siutsou@icranet.org}
\affiliation{ICRANet, Pescara, Italy\\
             65122, p-le della Repubblica, 10}
\affiliation{ICRA and Dip.\ di Fisica, Universita di Roma
\textquotedblleft Sapienza\textquotedblright, Roma, Italy\\
             00185, p-le A. Moro 5}
\affiliation{Stepanov Institute of Physics, NASB, Minsk, Belarus \\
              220072, Nezalezhnastsi Av., 68}
\author{Tomilchik L. M.}
\affiliation{Stepanov Institute of Physics, NASB, Minsk, Belarus \\
              220072, Nezalezhnastsi Av., 68, Tel.: +375-29-4040838}

\date{\today}

\pacs{04.25.Nx \and 04.80.Cc \and 04.50.Kd}

\begin{abstract}
From Doppler tracking data and data on circular motion of
astronomical objects we obtain a metric of the Pioneer Anomaly. The
metric resolves the issue of manifest absence of anomaly
acceleration in orbits of the outer planets and extra-Pluto objects
of the Solar system. However, it turns out that the energy-momentum
tensor of matter, which generates such a gravitational field in GR,
violates energy dominance conditions. At the same time the equation
of state derived from the energy-momentum tensor is that of dark
energy with $w=-1/3$. So the model proposed must be carefully
studied by "Grand-Fit" investigations.
\end{abstract}

\maketitle

\section{Introduction}
\label{intro} Spacecrafts Pioneer 10 and 11 were launched in the
early 1970's for the exploration of outer planets of the Solar
system (see the special issue of Science \textbf{183}, No. 4122, 25
January 1974, especially \cite{Soberman1974,Anderson1974}). After
the encounters with Jupiter and Saturn they followed hyperbolic
trajectories on leaving the Solar system. Because of rotational
spin-stabilization of these spacecrafts, which reduces the need for
manoeuvres, they represent unique experiments for testing our
understanding of celestial mechanics. The accuracy of acceleration
measurements for the Pioneer spacecrafts is about $10^{-10}$ m/s$^2$
\cite{PioneerMissionPlan,Scherer1997}.

During the flight spacecrafts were continuously tracked by Doppler
effect on retransmitted radio signals. Then data were fitted to
theoretical ones obtained in PPN approximation initially by ODP
program of JPL (JPL's Export Planetary Ephemeris DE405 was used for
planet motion).

But surprisingly above 10 a.u. of heliocentric distance the
systematical deviation of experimental and theoretical data was
found \cite{Pioneer1998}. This deviation can be described simply as a constant
acceleration towards the Sun with magnitude of about
$8\cdot10^{-10}$ m/s$^2$. This value is the same ---
within error limits --- for all the spaceships Pioneer 10 and 11,
Galileo and Ulysses and for all distances from the Sun \cite{Pioneer2001}.

This coincidence has been interpreted as a hint of the gravitational
--- metric --- origin of the acceleration. But at the same time there are no
signatures of such an acceleration in the orbits of outer planets
and other objects in the Solar system. Inclusion of such
acceleration leads to unavoidable deviations from the observed
planet positions \cite{Pioneer2006,Iorio2007}.

Many attempts to explain the anomaly were made during last 10 years.
Some of the recent work on this subject includes analyses of: the
thermal radiation of the Pioneers
\cite{ISI:000263816800017,ISI:000261214100012}, the gravitational
attraction by the Kuiper Belt
\cite{ISI:000238999400005,ISI:000239223200009,ISI:000232936700007},
the cosmological origin of the Anomaly
\cite{ISI:000255093800006,ISI:000242327900022,ISI:000255524300012,
ISI:000258636700088,ISI:000236229700005,ISI:000246464000017}, the
influence of multipole moments of the Sun
\cite{2005AIPC..758..129Q}, the clocks acceleration
\cite{ISI:000227551100006} and many proposes of modified gravity
\cite{ISI:000238120300014,ISI:000254557000003,ISI:000259935000020,
ISI:000257329300072} including even laboratory investigations on
very small acceleration dynamics \cite{ISI:000245691400011} and
interesting endeavors to constrain some parameters of modified
gravity theories by the known value of the Pioneer Anomaly
\cite{ISI:000248810500006,ISI:000262356900001}.

In the frame of metric theories of gravitation there is an
attractive possibility to explain the Pioneer Anomaly by metric
perturbation, preserving at the same time the character of planet
motion. It is possible because the Pioneers' trajectories are very
different from planet orbits: spacecrafts leave the Solar system
almost in a radial direction, while planets orbit the Sun almost
circularly. The potential possibility of such an approach was noted
independently in a series of manuscripts
\cite{2005CQGra..22.2135J,2006CQGra..23..777J,2006CQGra..23.7561J},
but authors of these works didn't analyze the origin of the metric.
Our approach is closer to that of Kjell Tangen \cite{Tangen2007},
but with more rigor because we don't neglect perturbation of the
space components of metric.

The goal of paper is to find the static space-time metric close to
Schwarzschild one, in which radial motion of test bodies shows the
Pioneer Anomaly, but circular motion doesn't. For this purpose in
section \ref{sec:1} we develop and discuss an algorithm of metric
determination from data on radial and circular motions. Metric
determination does not make use of Einstein equations so it is
applicable to any pure metric theory of gravity in terminology of
Will et. al. \cite{Will1993,Will2006}. Then we apply the method in
the case of the Pioneer Anomaly, starting from the Schwarzschild
space-time and recover the properties of matter forming such a
metric within GR (section \ref{sec:2}). Finally some concluding
remarks are made.

\section{Space-time determination from radial and circular motions}
\label{sec:1}

\subsection{Metric choice and time definition}
\label{inv}

We begin with the interval with 3 metric functions $\tau$, $\rho$
and $\sigma$ (for simplicity it was taken $c=1$)
\begin{gather}\label{metric}
ds^2=e^{\tau(r)}dt^2-e^{\rho(r)}dr^2-e^{\sigma(r)}r^2(d\theta^2+\cos^2\theta
d\varphi^2),
\end{gather}
and then find the gauge relation between them for maximal
simplification. The metric functions $\tau$, $\rho$ and $\sigma$
will be referred as time, radial and transverse metric function or
coefficient, respectively. We consider radial and circular motions
in such a space-time separately and find connections between metric
functions following from the known properties of the motion. But
first of all we must recall some time convention.

The metric (\ref{metric}) is written in the form that is consistent
with global clock synchronization, in fact $\partial/\partial t$
being timelike Killing vector. So in the approach proposed when the
cosmological effects is totally neglected and Solar system is
supposed to be placed in a space-time, that is Minkowskian at
spatial infinity, the coordinate time is the astronomical
ephemerides time \texttt{ET} (up to a multiplier). This is the usual
approximation used for PPN-ephemeride calculations.

There is a difficulty, because the perturbations required by the
Pioneer Anomaly grow with the radial distance, so the perturbed
space-time is not asymptotically flat. But it is not a big problem
because before the metric perturbations grow significantly the
space-time has a wide nearly-flat region in which we can choose
almost Minkowskian observers and coordinates. So further if we talk
about "spatial infinity", we mean this wide region, in the Solar
system scale the effects of difference between this approximation
and rigorous treatment are negligible. The same problem with the
same solution is arising in the PPN-approximation then we must place
the system not in Minkowskian background but in the cosmological
one. Additionally, as it can be shown, in PPN-approach the
cosmological effects, such as mutual acceleration of geodetically
moving bodies, have the second order in $H$. Therefore even while
the Pioneer Anomaly acceleration is nearly equal to $cH$, in the
framework of pure metric theories of gravitation there is no
possibility to link it to the cosmological expansion (early but
almost exhaustive analysis of the problem was made by R. C. Tolman
\cite[\S\S 153--156]{Tolman}, for the recent work on subject see the
articles mentioned in the Introduction and additionally
\cite{FahrSiewert2006,MizonyLachiezeRey2005,Licht2001} and
references within).

The radial coordinate rescaling $r\rightarrow f(r)$ changes all
metric functions, giving a possibility to imply gauge conditions on
the metric coefficients. But there are two invariants, i.e.
physically measurable quantities, which characterize the distance
from a given point to the center of space symmetry. Firstly it is an
atomic time rate in comparison with the coordinate time rate (or
atomic time rate on "spatial infinity"{}) $e^{\tau/2}$, and secondly
it is an area of a sphere
of points equidistant to the center of the space 
$4\pi r^2 e^{\sigma}$. So the numerical values of time and
transverse metric coefficients have clear physical meaning for a
given space-time point and only $\rho(r)$ can take arbitrary values.
Usual choices include $\rho\equiv\sigma$ corresponding to isotropic
coordinates of PPN-approximation and implicit on $\rho(r)$ relation
$\sigma\equiv0$ corresponding to Schwarzschild coordinates.

\subsection{Radial motion and its description by Doppler radio tracking}

Radial motion in the space-time is fully described by the energy
$g_{tt}\frac{dt}{ds}=e^{\tau(r)}u^0=k=const$ and the 4-velocity
length conservation
$e^{\tau(r)}u^{0^2}-e^{\rho(r)}u^{1^2}=\varepsilon,\ \varepsilon=0$
or 1 for electromagnetic waves and test bodies, respectively:
\begin{gather}\label{dt/dr}
 \frac{dt}{dr}=\frac{e^{\frac{\rho(r)-\tau(r)}2}}{\sqrt{1-\varepsilon
  e^{\tau(r)}/k^2}},
\end{gather}
the constant $k$ being connected with the velocity $v$ of the
spacecraft on "space infinity"{}
\begin{gather}
k^2=\frac{1}{1-v^2}.
\end{gather}
The relation (\ref{dt/dr}) can be integrated to give us $t(r)$
dependence which in turn can be inverted giving $r(t)$. But there is
a gauge freedom in the result because we can choose $\rho(r)$
freely. Moreover, this relation essentially involve both physical
$\tau$ and unphysical $\rho$.

Our goal is to determine the metric functions from observations.
Direct results of experimentation in astronomy and cosmology are the
measurements of externally originated signals received on the world
line of an observer. These signals have mostly electromagnetic
character and can originate from some external source (e. g. the
light emitted by the Sun and reflected by a planet) or be emitted by
the observer him/herself and then returned to him/her after some
interaction with outer bodies (as in case of radar measurements).
The latter case is
considered here as corresponding to the real situation \cite{Pioneer2001}.

The scheme of Doppler tracking used in the Pioneer experiments is
very simple conceptually: an electromagnetic signal, emitted from
the world line of the observer, is reflected back by the mirror on
the world line of the spacecraft and then compared with the initial
one again on the observer world line. To be more precise, the
monochromatic electromagnetic signal, obtained from the high
precision hydrogen maser \cite[subsection III.A]{Pioneer2001}, is
emitted by the antennae on Earth in the direction of the spacecraft.
This signal is detected by the spacecraft, amplified and reemitted
back to Earth, where the measured waveform of arrived signal is
compared with the emitted one to obtain the red shift of the signal
and the time of signal travel (the very detailed description of the
process can be found in \cite{Pioneer2001}).


Now we must describe the Doppler tracking and the signal time
arrival analysis in this space-time. In the geometric optics
approximation (which is applicable for the case considered) the
Doppler shift is governed simply by the ratio between scalar
products of the 4-velocity on world lines and the null wave vector
of the signal, parallel transported along the null geodesic line
between emitter and receiver:
\begin{gather}
 \frac{\nu_r}{\nu_e}=\frac{s_e}{s_r}=
\frac{\vec{\mathstrut u}_r\cdot \vec k_{r}}{\vec{\mathstrut
u}_e\cdot\vec k_{e}},
\end{gather}
where $\nu_r$ and $\nu_e$ are received and emitted frequencies,
measured by the standard atomic clocks,\\
$s_r$ and $s_e$ are proper times of one cycle of oscillation,\\
$u_r$ and $u_e$ are 4-velocities of receiver and emitter,\\
$k_r$ and $k_e$ are tangential null vectors (wave vector), parallel
transported along the path of the signal.

Atomic clock time deviations from ephemerides time along with all
known effects of Earth motion were taken into account during the
data processing (see \cite{Pioneer2001}), so for the description of
such a small deviation like the Pioneer Anomaly it is sufficient to
use the simple model, in which the emitter of the initial signal and
the receiver of the retranslated signal are fixed at constant
distances from the Sun on the line from the Sun to apparatus. The
signal is emitted from this "fixed" Earth at $r_0$ and $t-t_p$,
received by the spaceship at $r$ and $t$, amplified, exactly
retransmitted back to Earth and finally compared with the initial
frequency on the "fixed" Earth again at $r_0$ and $t+t_p$ ($t_p$ is
the time of signal propagation, the same for forward and backward
directions).

As it can be shown easily, the frequency $\nu_r$ received on Earth
is connected to the initially emitted $\nu_e$ as
\begin{gather}\label{nu_r}
 \nu_{r}=\nu_{e}\frac{1-\sqrt{1-e^{\tau(r)}/k^2}}{1+\sqrt{1-e^{\tau(r)}/k^2}}.
\end{gather}
This expression can be readily reduced to special relativistic one
in the case of $e^\tau\equiv1$.

As a relation between physical quantities only, this equation does
not involve arbitrary unphysical $\rho$ function. The problem of
gauge choice comes with the definition of $r(t)$: from (\ref{nu_r})
one can determine $e^{\tau(r(t))}$, but because the radial
coordinate $r$ (and consequently $r(t)$) is arbitrary, the radial
dependence of the time metric coefficient remains gauge dependent.
It is interesting also that this relation cannot be represented by a
power series in terms of small deviations of $e^{\tau(r)}$ and $k$
from 1. The transverse space metric coefficient $\sigma(r)$
naturally cannot be determined from the radial motion only.

So we come to an unavoidable alternative of \textit{a-priory} $r(t)$
definition or \textit{a-priory} imposing some gauge condition on
$\rho(r)$. In each case the remaining function is defined by the
experimental data, and our goal now is to find in which case the
process of metric restoration can be done without unnecessary
complications. In the next subsection we consider these
possibilities in some details and then show that the best choice is
the latter case, i.e. imposing a gauge.

\subsection{General formulae and coordinate choice}

While the time of signal arrival to the spaceship $t$ can be easily
determined as a half-sum of observed times of emitting $t_e$ and
receiving $t_r$ of the signal at the "fixed" Earth
\begin{gather}\label{t}
    t=\frac{t_r+t_e}2,
\end{gather}
the corresponding $r$ determination is not so trivial task. In
general we can measure only the time of signal travel from the
"fixed" Earth to the spaceship as a half-difference between observed
times of sending and receiving of the signal
\begin{gather}\label{tp}
 t_p=\int_{r_0}^r{e^{\frac{\rho(r)-\tau(r)}2}}\,dr=\frac{t_r-t_e}2.
\end{gather}
These are results of a different method of tracking
--- signal time arrival analysis, which in essence represents
integration of the Doppler data.

So to recover $\rho(r)$ from a given $r(t)$ we must firstly find
$\tau(r)$ by (\ref{nu_r}) from the observed redshift, and then solve
an integral equation above. On the other hand, to find $r(t)$ from a
given $\rho(r)$ it is sufficient to solve a non-integral equation
following from relation (\ref{dt/dr}) for the spacecraft motion
\begin{gather}
\int_{t_0}^t e^{\frac{\tau(r(t))}2}\sqrt{1-e^{\tau(r(t))}/k^2}\,dt=
\int_{r_0}^r {e^{\frac{\rho(r)}2}}\,dr,
\end{gather}
where $t_0$ is the time when the spacecraft leaves "fixed" Earth.
The left-hand side of the relation can be found totally from the
observed frequency shifts by (\ref{nu_r}), and the right-hand side
is a known function of $r$ with given $\rho(r)$.

Consequently maximal simplification of the problem is reached in the
case of \textit{a-priory} given radial metric function $\rho(r)$. It
is nonsense to define it dependent on still unknown time and
transverse metric functions, with one interesting exclusion: if
$\tau(r) \equiv \rho(r)$ then $r(t)$ can be recovered from
(\ref{tp}) simply as
\begin{gather}\label{r(t)}
    r=r_0+\frac{t_r-t_e}{2}.
\end{gather}
This choice of coordinates known as light or null coordinates is not
so usual as Schwarzschild ($\sigma\equiv0$) or isotropic
($\rho\equiv\sigma$) coordinates but it is the most suitable one for
the situation. Both abovementioned choices are especially
inappropriate here because they rely on transversal metric function
that does not reveal itself in pure radial motions.

\subsection{Circular motion}

We know that the near-circular motion of outer Solar system objects
(i.e. Neptune or Pluto) is unperturbed by the Pioneer Anomaly
acceleration \cite{Pioneer2006, Iorio2007}. This gives us a way to
determine the transversal metric coefficient and therefore to find
metric completely.

The angular velocity of circular motion ($\theta=0,\ \phi=\omega t$) in the
considered space-time is defined by the ratio of derivatives of time and
transversal space metric coefficients
\begin{gather}\label{omega}
 \omega^2(r)=\frac{(e^{\tau(r)})'}{(r^2e^{\sigma(r)})'}=
 \frac{(e^{\tau(r(t))}){}\dot{}}{(r^2e^{\sigma(r(t))}){}\dot{}},
\end{gather}
where as usual prime denotes differentiation with respect to radial
coordinate, and dot denotes differentiation in time. So there is no
dependence on the radial metric part at all. Moreover as
$\omega(r(t))$ and $(e^{\tau(r(t))}){}\dot{}$ is directly
observable, so the transverse metric coefficient
$r^2e^{\sigma(r(t))}$ can be obtained by a simple integration
without any notion of the radial metric function.

So again recalling the radial motion equation (\ref{dt/dr}) we
conclude that all gauge conditions for $\rho(r)$ involving
transverse metric coefficient are not convenient for treatment of
the Pioneer Anomaly, because all such conditions lead to coupling of
equations (\ref{nu_r}) and (\ref{omega}) which can be solved
independently otherwise.

\subsection{Final list of relations and concluding remarks}

So we work in light or null coordinates $\tau(r) \equiv \rho(r)$,
then:
\begin{gather}
\frac{dt}{dr}=\frac{1}{\sqrt{1-\varepsilon
e^{\tau(r)}/k^2}}.\\\intertext{The trajectory of the spacecraft $r(t)$ is
recovered simply from time of sending $t_e$ and arrival $t_r$ of signal}
 t_p=r-r_0 \quad\Rightarrow\quad t=\frac{t_r+t_e}{2},\qquad
r=r_0+\frac{t_r-t_e}{2},
\end{gather}
and the observed redshift of the signal
\begin{gather}
 z(t)=\frac{\nu_e-\nu_r}{\nu_e}=\frac{\Delta\nu}{\nu_e}
=\frac{2}
{(1-e^{\tau(r(t))}/k^2)^{-1/2}+1},
\end{gather}
can be immediately
transformed into time metric coefficient
\begin{gather}\label{tmc}
e^{\tau(r(t))}=k^2\left[1-\left(\frac{z(t)}{2-z(t)}\right)^2\right]=
4k^2\frac{(1-z)}{(2-z)^2}.
\end{gather}

The transverse space metric coefficient is defined by the dependence of
angular velocity $\omega$ on radial coordinate $r$
\begin{gather}\label{trans}
r^2e^{\sigma(r)}=r_0^2e^{\sigma(r_0)}-\int_{r_0}^r\frac{4k^2z(r)z'(r)}{
(2-z(r))^3 \omega^2(r)}dr,
\end{gather} which for the space-time of the Pioneer Anomaly should be the
same as in the Schwarzshild field (or very close to it). So basing
on the Schwarzshild metric one can find such perturbations of metric
coefficients, that the circular motion remains unperturbed, but the
radial one shows small deviation --- the Pioneer Anomaly.

It is worth noting that the sentence "the circular motion remains
unperturbed" denotes exactly the following: if, neglecting all
mutual planet disturbances, the period of circularly orbiting planet
will be measured in the units of ephemerides time and the distance
of planet from the baricenter of the Solar system will be found by
analyzing light propagation times on straight lines from the Sun
(null coordinates!) then the values of the period and the distance
will exactly be the same as needed for the 3rd Kepler law to hold
--- exactly as in the Schwarzshild field.

\section{The Pioneer Anomaly and its source in GR}
\label{sec:2}

\subsection{Schwarzschild space-time in light coordinates}

Radius $r$ in light coordinates of the Schwarzschild field is
related to the usual Schwarzschild radial coordinate $r_s$ as
\begin{gather}\label{rs}
 r=r_s+r_g\ln\left(\frac{r_s}{r_g}-1\right),\quad
 r_s=r_g\left(1+\mathop{W}\left(e^{\frac{r}{r_g}-1}\right)\right).
\end{gather}
The interval in null coordinates is as follows
\begin{gather}
\begin{split}ds^2=&\frac{\mathop{W}\left(e^{\frac{r}{r_g}-1}\right)}
{1+\mathop{W}\left(e^{\frac{r}{r_g}-1}\right)} (dt^2-dr^2)
\\&-r_g^2\left(1+\mathop{W}\left(e^{\frac{r}{
r_g}-1}\right)\right)^2(d\theta^2+\cos^2\theta
d\varphi^2),\end{split}\\\intertext{so that}
 e^{\tau(r)}=e^{\rho(r)}=\frac{\mathop{W}\left(e^{\frac{r}{r_g}-1}\right)}
{1+\mathop{W}\left(e^{\frac{r}{r_g}-1}\right)}=
r_g\mathop{W}\nolimits'_r\left(e^{\frac{r}{r_g}-1}\right),\\
e^{\sigma(r)}=\frac{r_g^2}{r^2}
\left(1+\mathop{W}\left(e^{\frac{r}{r_g}-1}\right)\right)^2,\\\intertext{where
$W(x)$ is the so called multiplicative logarithm or Lambert $W$ function}
W(x)e^{W(x)}=x.
\end{gather}

\subsection{The Pioneer Anomaly. Radial perturbation}

The Pioneer Anomaly is the linear in ephemerides time \texttt{ET}
deviation of the experimentally obtained frequency of received signal
$\nu_r$ from the "modelled" one $\nu_m$
\begin{equation}
\frac{d}{d\text{\texttt{ET}}}(\nu_r-\nu_m)=-\nu_e\frac{2a_P}{c},
\end{equation}
where $a_P\sim8\cdot10^{-10}\ \mbox{m/s}^2$ is the "unmodelled" acceleration
\cite{Pioneer2001}.
The ephemerides time 
coincides with time coordinate $t$ of the metric considered (as
described earlier). For the most part of range of the Pioneer
Anomaly found (from $\sim15$ to $\sim80$ a.u.) the deviation of the
Pioneers' orbits from pure radial motion is comparable to or even
below experimental uncertainty in the acceleration: it can be
estimated roughly as a ratio of the semi-major axis $a$ absolute
value to heliocentric distance $r$
\begin{equation}
    \frac{|a|}{r}\lesssim\frac{10^9 \text{km}}{40\cdot150\cdot10^6
    \text{km}}\simeq17 \%,
\end{equation}
while experimental error in the acceleration is $1.33/8.74\simeq15
\%$ (see Appendix of \cite{Pioneer2001}). It should be noted that
this uncertainty prevents Anderson et. al. from determining the
direction of the acceleration: to Earth or to the Sun (see beginning
of section VII and especially note 73 of \cite{Pioneer2001}).

The modelled frequency and velocity of the spacecraft are
\begin{gather}
\nu_{m}=\nu_0\;\frac{1+\frac1{W(e^{\frac{r}{r_g}-1})}}{1-v^2}
\left(1-\sqrt{1-\frac{1-v^2}{1+\frac1{W(e^{\frac{r}{r_g}-1})}}}\right)^2,
\\[-0.5ex]
\left(\frac{dr}{dt}\right)_m=v_m(r)=
\sqrt{\frac{v^2+1/W(e^{\frac{r}{r_g}-1})}{1+1/W(e^{\frac{r}{r_g}-1})}},
\end{gather}
so accurately expanding the expression (\ref{tmc}) for the time
metric coefficient from the red shift $z(r(t))$, one can find that
given the accuracy of the Pioneer Anomaly measurements one can
simply use the relation
\begin{gather}\label{PAmain}
 \delta e^{\tau(r)}\simeq-\frac{z(r)\;\delta z(r)}2=a_P\;z(r)\Delta t(r),
\end{gather}
where $\Delta t=t(r)-t_0$ is the time from the start of the Pioneer
Anomaly (we suppose that for $r<r_0$ the metric coincides with the
Schwarzschild one, so before $t_0=t(r_0)$ there is no anomalous
acceleration), $\delta z(r)$ is a deviation of observed red shift
from the modelled one.

\begin{figure}
\centerline{\includegraphics[scale=1.00]{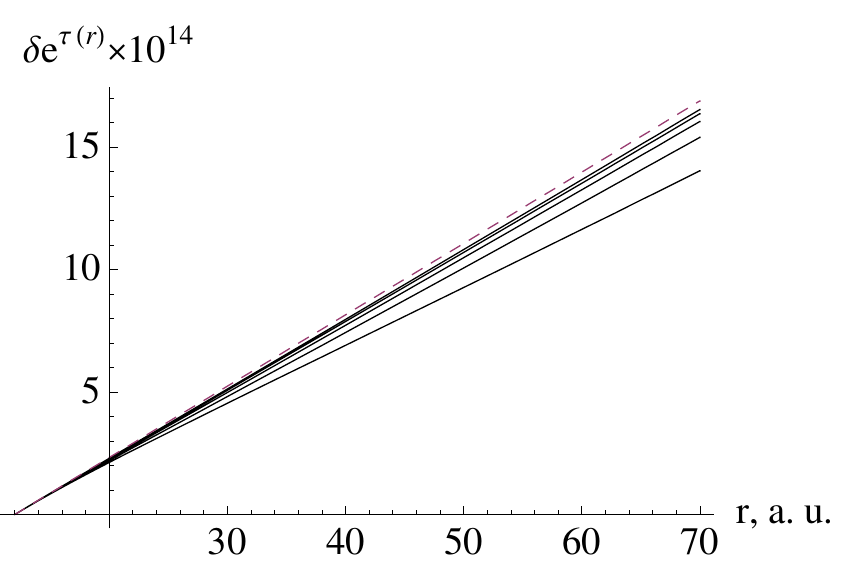}}
\caption{\label{deltatau}Metric perturbation $\delta e^{\tau(r)}$ of
the Pioneer Anomaly for $v$ from 5~km/s to 50~km/s in 5~km/s steps
(solid lines from bottom to top) for the metric matching the
Schwarzschild metric at 12 a.\;u., compared to the "naive"
post-Newtonian one (dashed line)}
\end{figure}

In the first approximation $t(r)$ dependence can be replaced with
the "modelled" time, which to the accuracy of the measurements is
the same as in the Newtonian case
\begin{multline}
 t_m(r)=t(r_0)+\int_{r_0}^{r}\frac{dr}{\dot r}\simeq\\
\simeq t_0+\frac{r}{v^2}\sqrt{v^2+\frac{r_g}{r}}-\frac{r_g}{v^3}
\sinh^{-1}\left(\sqrt{\frac{r}{r_g}} v\right).
\end{multline}
Finally inserting this and modelled $z(r)$ into the equation
(\ref{PAmain}) we arrive to the perturbation of the time metric
coefficient
\begin{multline}
 \delta e^{\tau(r)} =
2a_P\Biggl(r-C \sqrt{v^2+\frac{r_g}{r}}+\\+\frac{r_g}{v^2}\left[
1-\sqrt{1+\frac{r_g}{r\;v^2}}
\sinh^{-1}\left(\sqrt{\frac{r}{r_g}}v\right) \right] \Biggr),
\end{multline}
where $C$ is a constant which can be determined from $r_0$ and $v$.
The perturbation appears to be non-linear in $r$, but for the
Pioneer 10/11 parameters the deviation from linearity is buried deep
in the experimental errors. It is illustrated by the figure
\ref{deltatau}, which shows the deviation of the time metric
coefficient compared to the "naive" post-Newtonian approach, where
one simply adds to the gravitational potential $\Phi(r)$ a term
linear in radius and use $e^{\tau(r)}\simeq1+\Phi(r)$. As we can
see, the difference is mainly in the slope, all the graphs are
nearly linear. Moreover, the relative value of the deviation from
linearity is decreasing with radial distance. So the linear
approximation $\delta e^{\tau(r)} \simeq 2\eta a_P (r-r_0)$ is
sufficient for the Pioneer Anomaly explanation. The only difference
between this more accurate result and the "naive" post-Newtonian one
is the presence of $\eta$, which is always less then 1 (see table
\ref{eta}).

\begin{table*}
 \caption{The quantity $1-\eta$ of the best linear approximations $\delta
e^{\tau(r)}\simeq 2\eta a_P (r-r_0)$ on the interval $r_0\leq r \leq
70$ a.\,u. for different velocities $v$ and metric matching
distances $r_0{}^{*}$ } \label{eta}
\begin{tabular}{@{\hspace{4mm}}c@{\hspace{6mm}}c@{\hspace{4mm}}c%
@{\hspace{4mm}}c@{\hspace{4mm}}c@{\hspace{4mm}}c@{\hspace{4mm}}c%
@{\hspace{4mm}}c@{\hspace{4mm}}c@{\hspace{4mm}}c@{\hspace{4mm}}c@{\hspace{4mm}}}
\hline\noalign{\smallskip}
 \multirow{2}*{$r_0$, a.\;u.}&\multicolumn{10}{c}{$v$, km/s\strut}\\
 \noalign{\smallskip}\cline{2-11}\rule[1ex]{0pt}{2ex}
&5&10&15&20&25&30&35&40&45&50\\\noalign { \smallskip}\hline\noalign{\smallskip}
 10 & 0.175 & 0.098 & 0.057 & 0.036 & 0.025 &
0.018 & 0.013 & 0.010 & 0.008 & 0.007 \\\noalign{\smallskip}
 15 & 0.146 & 0.077 & 0.044 & 0.027 & 0.018 & 0.013 &
0.010 & 0.008 & 0.006 & 0.005 \\\noalign{\smallskip}
 20 & 0.122 & 0.062 & 0.034 & 0.021 & 0.014 & 0.010 &
0.007 & 0.006 & 0.005 &
0.004\\\noalign{\smallskip}\hline\noalign{\smallskip}
\multicolumn{11}{p{0.7\textwidth}}{\small$^*$ $r_0$ is the radial
coordinate at which the metric coefficients coincides with the
Schwarzschild ones.}
\end{tabular}
\end{table*}

\subsection{Transversal perturbation leaving planet orbits unchanged.
Necessity of the central source}

Now we assume that in the perturbed space-time planets orbit with
the same periods as in the unperturbed Schwarzschild solution, so
that there is no signatures of the Pioneer acceleration in the
orbits of planets. So the angular velocity dependence $\omega(r)$
must be the same as in the Schwarzschild case (see, e. g.,
eq.~(25.40) of \cite{MTW})
\begin{gather}
 \omega_s^2(r)=\frac{{e^{\tau(r)}}'}{(r^2e^{\sigma(r)})'}=
\frac{1}{2r_g^2\left(1+\mathop{W}\left(e^{\frac{r}{r_g}-1}\right)\right)^3}=
\frac{r_g}{2r_s^3},
\end{gather}
and the perturbation in $e^\tau$ must lead to such a perturbation in
$r^2e^\sigma$ that $\omega^2(r)$ remains invariant. In the perturbed
space-time by the general formula (\ref{omega}) 
we have from simple mathematics
\begin{multline}
 \omega_s^2(r)=\frac{{e^{\tau(r)}}'}{(r^2e^{\sigma(r)})'}=\omega^2(r)=
\frac{(e^{\tau(r)}+\delta e^{\tau(r)})'}{(r^2e^{\sigma(r)}+ \delta
(r^2e^{\sigma(r)}))'}=\\=\frac{(e^{\tau(r)})'+(\delta
e^{\tau(r)})'}{(r^2e^{\sigma(r)})'+(\delta (r^2e^{\sigma(r)}))'}=
\frac{(\delta e^{\tau(r)})'}{(\delta(r^2e^{\sigma(r)}))'}.
\end{multline}
So the perturbation of the transverse metric coefficient is
\begin{gather}
\delta (r^2e^{\sigma(r)})=r^2 \delta e^{\sigma(r)}=
\int_{r_0}^r \omega^{-2}(r)\delta({e^{\tau(r)}})'\,dr,
\end{gather}
and finally
\begin{multline}
 \delta e^{\sigma(r)}=
\frac{4a_P \eta\, r_g^2}{r^2} \int_{r_0}^r
\left(1+\mathop{W}\left(e^{\frac{r}{r_g}-1}\right)\right)^3\,dr
\simeq\\\simeq\frac{4a_P \eta\, r_g^2}{r^2}
\int_{r_0}^r\left(\frac{r}{r_g}\right)^3 dr= \frac{4a_P \eta
(r^4-r_0^4)}{r^2 r_g}.
\end{multline}
In the first approximation the perturbation of $r^2e^{\sigma(r)}$ grows
quartically in radius.

Note the gravitational radius of the source $r_g$ in the answer. So
\emph{the effect of the Pioneer Anomaly can be reproduced only by
perturbations of Schwarzschild space-time and not Minkowski one}. So
the gravitational explanation of the Pioneer Anomaly can be obtained
without the equivalence principle violation required by various
authors \cite{Pioneer2006,Iorio2007,Tangen2007}.

\subsection{Matter corresponding to the obtained metric in General Relativity}

One can try to find the gravitational field theory, that gives
equations of the gravitational field allowing the solution found for
the weak field of a point mass. But we think that it is not very
promising because there are no reliable experimental evidence in
favor of any gravitational theory other than General Relativity.

Instead we find the properties of matter surrounding the point mass
in GR that can generate the obtained metric. Because in the scale of
Solar system experiments the influence of cosmological constant is
negligible, for the determination of matter one can use the Einstein
equations of the form
\begin{equation}
 G_{ij} = R_{ij} - \frac12 R g_{ij} = \kappa T_{ij},\qquad
 \kappa=\frac{8\pi G}{c^4}.
\end{equation}
Using for simplicity the metric in the form
\begin{equation}
 ds^2=e^{\tau(r)}dt^2-e^{\rho(r)}dr^2-e^{\sigma(r)}(d\theta^2+\sin^2\theta
d\phi^2),
\end{equation}
one arrives at the Einstein tensor
\begin{gather}
G_{ij}=\frac{e^{-\rho}}{4}\Bigl(\lambda_t T_i\otimes T_j-\lambda_s
S_i\otimes S_j-\lambda g_{ij}\Bigr),\\
 S_i=\Bigl\{0,e^{\frac{\rho}{2}},0,0\Bigr\},\qquad
 T_i=\Bigl\{e^{\frac{\tau}{2}},0,0,0\Bigr\},\\
 -S_i S^i=T_i T^i=1,\qquad S^i T_i=0,\\
 \lambda_t=4 e^{\rho-\sigma}+\left(\rho'-2 \sigma'-\tau'\right)
\left(\sigma'-\tau'\right)+2 \left(\tau''-\sigma''\right),\\
\lambda_s=4 e^{\rho-\sigma}-\tau'\left(\sigma'-\tau'\right)-\rho'
\left(\sigma'+\tau'\right)+2 \left(\tau''+\sigma''\right),\\
\lambda=\sigma'^2+\sigma'\tau'+\tau'^2-\rho'\left(\sigma'+\tau'\right)+
2\left(\tau''+\sigma''\right).
\end{gather}

Expanding to the first power of $a_P$ one obtains
\begin{gather}
 \lambda_t=-96\, \frac{a_P \eta}{r_g},\quad
 \lambda_s=-32\, \frac{a_P \eta}{r_g}\,\frac{r_0^4}{r^4},\nonumber\\
 \lambda=16\,\frac{a_P \eta}{r_g}\left(3-\frac{r_0^4}{r^4}\right).
\end{gather}
The algebraic type of the energy-momentum tensor at spatial infinity
is that of an ideal fluid (by $\lambda_s\rightarrow0$) with constant
positive pressure
\begin{gather}
 p=\frac{e^{-\tau}}{4\kappa}\lambda\rightarrow 12\,
\frac{a_P\eta}{\kappa r_g}>0,\\
\intertext{but \emph{negative energy density}}
 {\rho=\frac{e^{-\tau}}{4\kappa}(\lambda_t-\lambda) \rightarrow
-36\, \frac{a_P \eta}{\kappa r_g}<0.}
\end{gather}
It is worth noting that relation between $p$ and $\rho$ is as for an
ultrarelativistic fluid except for the sign: instead of $p=\rho/3$
one has asymptotically $p=-\rho/3$. This is a typical equation of
state of dark energy with parameter $w=-1/3$. It is interesting that
such a fluid does not change the cosmological dynamics of
Friedman-Lema\^{i}tre universe (see, e.g.,
\cite[III.E]{PeeblesRatra2003}).

\section{Conclusions}

In this work we show that it is possible to perturb time metric
coefficient of the Schwarzschild space-time in such a way that the
Pioneer Anomaly is reproduced. Moreover, because planet motion is
governed also by anther component of the metric, we can tune it so
that circular orbits is not disturbed. This result applies in any
pure metric theory of gravitation where test bodies follow geodetics
of the metric. It is deduced that the perturbation of the time
metric coefficient can be taken linear in $r$ without contradiction
with the accuracy of experimental data obtained up to now.

Assuming the validity of the General Relativity, we find out the
energy-momentum tensor generating the metric obtained. The tensor
corresponds to an ideal fluid, however having negative energy
density. It is interesting to note that an exact static solution
with spherical symmetry is known for the "fluid" with $3p+\rho=0$
\cite[\S 8.5]{StanjukovichMel'nikov1983}. This "fluid" does not
interact with the ordinary matter besides its gravitational
influence on the metric, so it much like WIMPs or scalar field of
gravitational theories of Brans-Dicke type. Thus the ordinary matter
including spacecrafts and planets is moving geodesically.

Naturally the found "fluid" does change the planet orbits (if they
are not strictly circular) and light rays paths. The model proposed
must be carefully studied in view of the "Grand-Fit" investigations
\cite{Pitjeva2005, 2008AIPC..977..254S}, but direct measurements
from the planned missions for testing General Relativity in space
are preferable \cite{2007arXiv0711.0304W, 2005gr.qc.....6104L,
2005gr.qc.....6139T}. The absolute value of effects for the
perturbation found as well as for exact solution of Stanjukovich and
Ivanov will be studied in the forthcoming paper.

The analysis presented in this paper can encourage someone to find
out which of the known alternatives to GR can reproduce the metric
found or to invent some new theory which can do it. It is possible,
but in our opinion the Pioneer Anomaly has some simple explanation
by conventional and non-gravitational physics, which is not found
yet. So the question of deep theoretical grounds for the existence
of the "fluid" in GR or of development of some new theory of
gravitation based on the metric obtained is not in the scope of our
article. Instead we point out that negative energy density of the
"fluid" is in a direct contradiction with the properties of
conventional matter. One interesting possibility for the "fluid" is
dark energy, which has the right equation of state. Therefore we
suppose that at present the metric (gravitational) origin of the
Pioneer Anomaly cannot be ruled out.

\begin{acknowledgements}{The authors thank participants of the
Theoretical Physics Laboratory seminar for helpful discussions and
fruitful remarks, and also acknowledge the creator of the RGTC
package --- S. Bonanos. One of us (SIA) thanks G. Vereshchagin for
the help in preparation of the present version of manuscript.}
\end{acknowledgements}


\end{document}